# BPR: Bayesian Personalized Ranking from Implicit Feedback

**Steffen Rendle, Christoph Freudenthaler, Zeno Gantner and Lars Schmidt-Thieme**
{srendle, freudenthaler, gantner, schmidt-thieme}@ismll.de
Machine Learning Lab, University of Hildesheim
Marienburger Platz 22, 31141 Hildesheim, Germany

## Abstract

Item recommendation is the task of predicting a personalized ranking on a set of items (e.g. websites, movies, products). In this paper, we investigate the most common scenario with implicit feedback (e.g. clicks, purchases). There are many methods for item recommendation from implicit feedback like matrix factorization (MF) or adaptive k-nearest-neighbor (kNN). Even though these methods are designed for the item prediction task of personalized ranking, none of them is directly optimized for ranking. In this paper we present a generic optimization criterion BPR-Opt for personalized ranking that is the maximum posterior estimator derived from a Bayesian analysis of the problem. We also provide a generic learning algorithm for optimizing models with respect to BPR-Opt. The learning method is based on stochastic gradient descent with bootstrap sampling. We show how to apply our method to two state-of-the-art recommender models: matrix factorization and adaptive kNN. Our experiments indicate that for the task of personalized ranking our optimization method outperforms the standard learning techniques for MF and kNN. The results show the importance of optimizing models for the right criterion.

## 1 Introduction

Recommending content is an important task in many information systems. For example online shopping websites like Amazon give each customer personalized recommendations of products that the user might be interested in. Other examples are video portals like YouTube that recommend movies to customers. Personalization is attractive both for content providers, who can increase sales or views, and for customers, who can find interesting content more easily. In this paper, we focus on item recommendation. The task of item recommendation is to create a user-specific ranking for a set of items. Preferences of users about items are learned from the user's past interaction with the system – e.g. his buying history, viewing history, etc.

Recommender systems are an active topic of research. Most recent work is on scenarios where users provide explicit feedback, e.g. in terms of ratings. Nevertheless, in real-world scenarios most feedback is not explicit but implicit. Implicit feedback is tracked automatically, like monitoring clicks, view times, purchases, etc. Thus it is much easier to collect, because the user has not to express his taste explicitly. In fact implicit feedback is already available in almost any information system – e.g. web servers record any page access in log files.

In this paper we present a generic method for learning models for personalized ranking. The contributions of this work are:

1. We present the generic optimization criterion BPR-Opt derived from the maximum posterior estimator for optimal personalized ranking. We show the analogies of BPR-Opt to maximization of the area under ROC curve.

2. For maximizing BPR-Opt, we propose the generic learning algorithm LearnBPR that is based on stochastic gradient descent with bootstrap sampling of training triples. We show that our algorithm is superior to standard gradient descent techniques for optimizing w.r.t. BPR-Opt.

3. We show how to apply LearnBPR to two state-of-the-art recommender model classes.

4. Our experiments empirically show that for the task of of personalized ranking, learning a model with BPR outperforms other learning methods.

## 2 Related Work

The most popular model for recommender systems is k-nearest neighbor (kNN) collaborative filtering [2]. Traditionally the similarity matrix of kNN is computed by heuristics – e.g. the Pearson correlation – but in recent work [8] the similarity matrix is treated as model parameters and is learned specifically for the task. Recently, matrix factorization (MF) has become very popular in recommender systems both for implicit and explicit feedback. In early work [13] singular value decomposition (SVD) has been proposed to learn the feature matrices. MF models learned by SVD have shown to be very prone to overfitting. Thus regularized learning methods have been proposed. For item prediction Hu et al. [5] and Pan et al. [10] propose a regularized least-square optimization with case weights (WR-MF). The case weights can be used to reduce the impact of negative examples. Hofmann [4] proposes a probabilistic latent semantic model for item recommendation. Schmidt-Thieme [14] converts the problem into a multi-class problem and solves it with a set of binary classifiers. Even though all the work on item prediction discussed above is evaluated on personalized ranking datasets, none of these methods directly optimizes its model parameters for ranking. Instead they optimize to predict if an item is selected by a user or not. In our work we derive an optimization criterion for personalized ranking that is based on pairs of items (i.e. the user-specific order of two items). We will show how state-of-the-art models like MF or adaptive kNN can be optimized with respect to this criterion to provide better ranking quality than with usual learning methods. A detailed discussion of the relationship between our approach and the WR-MF approach of Hu et al. [5] and Pan et al. [10] as well as maximum margin matrix factorization [15] can be found in Section 5. In Section 4.1.1, we will also discuss the relations of our optimization criterion to AUC optimization like in [3].

In this paper, we focus on offline learning of the model parameters. Extending the learning method to online learning scenarios – e.g. a new user is added and his history increases from 0 to 1, 2, ... feedback events – has already been studied for MF for the related task of rating prediction [11]. The same fold-in strategy can be used for BPR.

There is also related work on learning to rank with non-collaborative models. One direction is to model distributions on permutations [7, 6]. Burges et al. [1] optimize a neural network model for ranking using gradient descent. All these approaches learn only one ranking – i.e. they are non-personalized. In contrast to this, our models are collaborative models that learn personalized rankings, i.e. one individual ranking per user. In our evaluation, we show empirically that in typical recommender settings our personalized BPR model outperforms even the theoretical upper bound for non-personalized ranking.

## 3 Personalized Ranking

The task of personalized ranking is to provide a user with a ranked list of items. This is also called item recommendation. An example is an online shop that wants to recommend a personalized ranked list of items that the user might want to buy. In this paper we investigate scenarios where the ranking has to be inferred from the implicit behavior (e.g. purchases in the past) of the user. Interesting about implicit feedback systems is that only positive observations are available. The non-observed user-item pairs – e.g. a user has not bought an item yet – are a mixture of real negative feedback (the user is not interested in buying the item) and missing values (the user might want to buy the item in the future).

### 3.1 Formalization

Let $U$ be the set of all users and $I$ the set of all items. In our scenario implicit feedback $S \subseteq U \times I$ is available (see left side of Figure 1). Examples for such feedback are purchases in an online shop, views in a video portal or clicks on a website. The task of the recommender system is now to provide the user with a personalized total ranking $>_u \subset I^2$ of all items, where $>_u$ has to meet the properties of a total order:

$$\forall i, j \in I : i \neq j \Rightarrow i >_u j \vee j >_u i \quad \text{(totality)}$$
$$\forall i, j \in I : i >_u j \wedge j >_u i \Rightarrow i = j \quad \text{(antisymmetry)}$$
$$\forall i, j, k \in I : i >_u j \wedge j >_u k \Rightarrow i >_u k \quad \text{(transitivity)}$$

For convenience we also define:

$$I_u^+ := \{i \in I : (u, i) \in S\}$$
$$U_i^+ := \{u \in U : (u, i) \in S\}$$

### 3.2 Analysis of the problem setting

As we have indicated before, in implicit feedback systems only positive classes are observed. The remaining data is a mixture of actually negative and missing values. The most common approach for coping with the missing value problem is to ignore all of them but then typical machine learning models are unable to learn anything, because they cannot distinguish between the two levels anymore.

The usual approach for item recommenders is to predict a personalized score $\hat{x}_{ui}$ for an item that reflects the preference of the user for the item. Then the items are ranked by sorting them according to that score. Machine learning approaches for item recommenders [5, 10] typically create the training data from $S$ by giving pairs $(u, i) \in S$ a positive class label and all other combinations in $(U \times I) \setminus S$ a negative one (see Figure 1). Then a model is fitted to this data. That means the model is optimized to predict the value 1 for elements in $S$ and 0 for the rest. The problem with this approach is that all elements the model should rank in the future $((U \times I) \setminus S)$ are presented to the learning algorithm as negative feedback during training. That means a model with enough expressiveness (that can fit the training data exactly) cannot rank at all as it predicts only 0s. The only reason why such machine learning methods can predict rankings are strategies to prevent overfitting, like regularization.

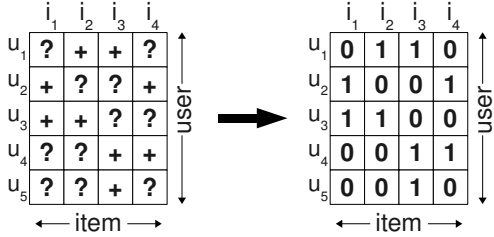

Figure 1: On the left side, the observed data $S$ is shown. Learning directly from $S$ is not feasible as only positive feedback is observed. Usually negative data is generated by filling the matrix with 0 values.

We use a different approach by using item pairs as training data and optimize for correctly ranking item pairs instead of scoring single items as this better represents the problem than just replacing missing values with negative ones. From $S$ we try to reconstruct for each user parts of $>_u$. If an item $i$ has been viewed by user $u$ – i.e. $(u, i) \in S$ – then we assume that the user prefers this item over all other non-observed items. E.g. in Figure 2 user $u_1$ has viewed item $i_2$ but not item $i_1$, so we assume that this user prefers item $i_2$ over $i_1$: $i_2 >_u i_1$. For items that have both been seen by a user, we cannot infer any preference. The same is true for two items that a user has not seen yet (e.g. item $i_1$ and $i_4$ for user $u_1$). To formalize this we create training data $D_S : U \times I \times I$ by:

$$D_S := \{(u, i, j) | i \in I_u^+ \wedge j \in I \setminus I_u^+\}$$

The semantics of $(u, i, j) \in D_S$ is that user $u$ is assumed to prefer $i$ over $j$. As $>_u$ is antisymmetric, the negative cases are regarded implicitly.

Our approach has two advantages:

1. Our training data consists of both positive and

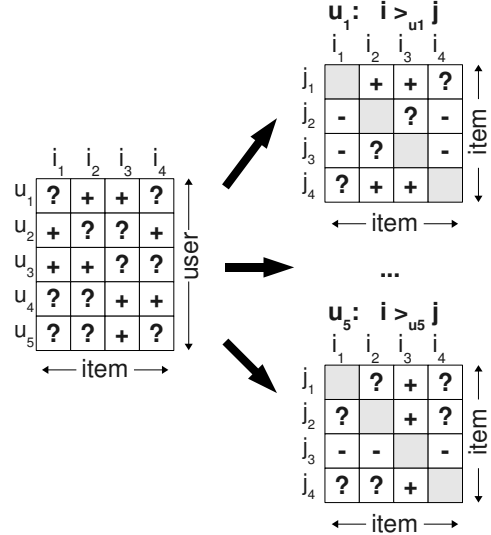

Figure 2: On the left side, the observed data $S$ is shown. Our approach creates user specific pairwise preferences $i >_u j$ between a pair of items. On the right side, plus (+) indicates that a user prefers item $i$ over item $j$; minus (−) indicates that he prefers $j$ over $i$.

negative pairs and missing values. The missing values between two non-observed items are exactly the item pairs that have to be ranked in the future. That means from a pairwise point of view the training data $D_S$ and the test data is disjoint.

2. The training data is created for the actual objective of ranking, i.e. the observed subset $D_S$ of $>_u$ is used as training data.

## 4 Bayesian Personalized Ranking (BPR)

In this section we derive a generic method for solving the personalized ranking task. It consists of the general optimization criterion for personalized ranking, BPR-OPT, which will be derived by a Bayesian analysis of the problem using the likelihood function for $p(i >_u j | \Theta)$ and the prior probability for the model parameter $p(\Theta)$. We show the analogies to the ranking statistic AUC (area under the ROC curve). For learning models with respect to BPR-OPT, we propose the algorithm LEARNBPR. Finally, we show how BPR-OPT and LEARNBPR can be applied to two state-of-the-art recommender algorithms, matrix factorization and adaptive kNN. Optimized with BPR these models are able to generate better rankings than with the usual training methods.

### 4.1 BPR Optimization Criterion

The Bayesian formulation of finding the correct personalized ranking for all items $i \in I$ is to maximize the following posterior probability where $\Theta$ represents the parameter vector of an arbitrary model class (e.g. matrix factorization).

$$p(\Theta|>_u) \propto p(>_u |\Theta)\, p(\Theta)$$

Here, $>_u$ is the desired but latent preference structure for user $u$. All users are presumed to act independently of each other. We also assume the ordering of each pair of items $(i, j)$ for a specific user is independent of the ordering of every other pair. Hence, the above user-specific likelihood function $p(>_u |\Theta)$ can first be rewritten as a product of single densities and second be combined for all users $u \in U$.

$$\prod_{u \in U} p(>_u |\Theta) = \prod_{(u,i,j) \in U \times I \times I} p(i >_u j|\Theta)^{\delta((u,i,j) \in D_S)} \cdot (1 - p(i >_u j|\Theta))^{\delta((u,j,i) \notin D_S)}$$

where $\delta$ is the indicator function:

$$\delta(b) := \begin{cases} 1 & \text{if } b \text{ is true,} \\ 0 & \text{else} \end{cases}$$

Due to the totality and antisymmetry of a sound pairwise ordering scheme the above formula can be simplified to:

$$\prod_{u \in U} p(>_u |\Theta) = \prod_{(u,i,j) \in D_S} p(i >_u j|\Theta)$$

So far it is generally not guaranteed to get a personalized total order. In order to establish this, the already mentioned sound properties (totality, antisymmetry and transitivity) need to be fulfilled. To do so, we define the individual probability that a user really prefers item $i$ to item $j$ as:

$$p(i >_u j|\Theta) := \sigma(\hat{x}_{uij}(\Theta))$$

where $\sigma$ is the logistic sigmoid:

$$\sigma(x) := \frac{1}{1 + e^{-x}}$$

Here $\hat{x}_{uij}(\Theta)$ is an arbitrary real-valued function of the model parameter vector $\Theta$ which captures the special relationship between user $u$, item $i$ and item $j$. In other words, our generic framework delegates the task of modeling the relationship between $u$, $i$ and $j$ to an underlying model class like matrix factorization or adaptive kNN, which are in charge of estimating $\hat{x}_{uij}(\Theta)$. Hence, it becomes feasible to statistically model a personalized total order $>_u$. For convenience, in the following we will skip the argument $\Theta$ from $\hat{x}_{uij}$.

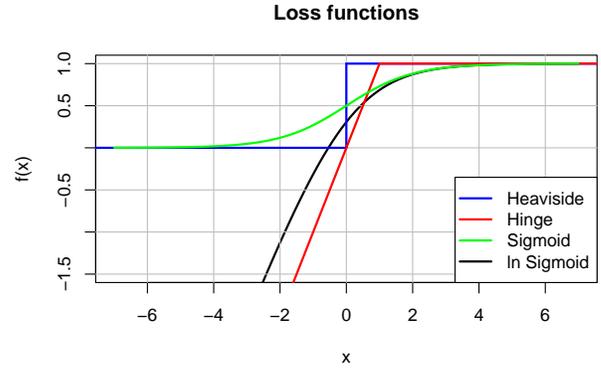

Figure 3: Loss functions for optimizing the AUC. The non-differentiable Heaviside $H(x)$ is often approximated by the sigmoid $\sigma(x)$. Our MLE derivation suggests to use $\ln \sigma(x)$ instead.

So far, we have only discussed the likelihood function. In order to complete the Bayesian modeling approach of the personalized ranking task, we introduce a general prior density $p(\Theta)$ which is a normal distribution with zero mean and variance-covariance matrix $\Sigma_\Theta$.

$$p(\Theta) \sim N(0, \Sigma_\Theta)$$

In the following, to reduce the number of unknown hyperparameters we set $\Sigma_\Theta = \lambda_\Theta I$. Now we can formulate the maximum posterior estimator to derive our generic optimization criterion for personalized ranking BPR-OPT.

$$\begin{aligned}\text{BPR-OPT} &:= \ln p(\Theta|>_u) \\ &= \ln p(>_u |\Theta)\, p(\Theta) \\ &= \ln \prod_{(u,i,j) \in D_S} \sigma(\hat{x}_{uij})\, p(\Theta) \\ &= \sum_{(u,i,j) \in D_S} \ln \sigma(\hat{x}_{uij}) + \ln p(\Theta) \\ &= \sum_{(u,i,j) \in D_S} \ln \sigma(\hat{x}_{uij}) - \lambda_\Theta ||\Theta||^2\end{aligned}$$

where $\lambda_\Theta$ are model specific regularization parameters.

#### 4.1.1 Analogies to AUC optimization

With this formulation of the Bayesian Personalized Ranking (BPR) scheme, it is now easy to grasp the analogy between BPR and AUC. The AUC per user is

usually defined as:

$$\text{AUC}(u) := \frac{1}{|I_u^+||I \setminus I_u^+|} \sum_{i \in I_u^+} \sum_{j \in |I \setminus I_u^+|} \delta(\hat{x}_{uij} > 0)$$

Hence the average AUC is:

$$\text{AUC} := \frac{1}{|U|} \sum_{u \in U} AUC(u)$$

With our notation of $D_S$ this can be written as:

$$\text{AUC}(u) = \sum_{(u,i,j) \in D_S} z_u \, \delta(\hat{x}_{uij} > 0) \qquad (1)$$

where $z_u$ is the normalizing constant:

$$z_u = \frac{1}{|U| \, |I_u^+| \, |I \setminus I_u^+|}$$

The analogy between (1) and BPR-OPT is obvious. Besides the normalizing constant $z_u$ they only differ in the loss function. The AUC uses the non-differentiable loss $\delta(x > 0)$ which is identical to the Heaviside function:

$$\delta(x > 0) = H(x) := \begin{cases} 1, & x > 0 \\ 0, & \text{else} \end{cases}$$

Instead we use the differentiable loss $\ln \sigma(x)$. It is common practice to replace the non-differentiable Heaviside function when optimizing for AUC [3]. Often the choice of the substitution is heuristic and a similarly shaped function like $\sigma$ is used (see figure 3). In this paper, we have derived the alternative substitution $\ln \sigma(x)$ that is motivated by the MLE.

### 4.2 BPR Learning Algorithm

In the last section we have derived an optimization criterion for personalized ranking. As the criterion is differentiable, gradient descent based algorithms are an obvious choice for maximization. But as we will see, standard gradient descent is not the right choice for our problem. To solve this issue we propose LEARNBPR, a stochastic gradient-descent algorithm based on bootstrap sampling of training triples (see figure 4).

First of all the gradient of BPR-OPT with respect to the model parameters is:

$$\frac{\partial \text{BPR-OPT}}{\partial \Theta} = \sum_{(u,i,j) \in D_S} \frac{\partial}{\partial \Theta} \ln \sigma(\hat{x}_{uij}) - \lambda_\Theta \frac{\partial}{\partial \Theta} ||\Theta||^2$$

$$\propto \sum_{(u,i,j) \in D_S} \frac{-e^{-\hat{x}_{uij}}}{1 + e^{-\hat{x}_{uij}}} \cdot \frac{\partial}{\partial \Theta} \hat{x}_{uij} - \lambda_\Theta \Theta$$

1: **procedure** LEARNBPR($D_S, \Theta$)
2:    initialize $\Theta$
3:    **repeat**
4:       draw $(u,i,j)$ from $D_S$
5:       $\Theta \leftarrow \Theta + \alpha \left( \frac{e^{-\hat{x}_{uij}}}{1+e^{-\hat{x}_{uij}}} \cdot \frac{\partial}{\partial \Theta} \hat{x}_{uij} + \lambda_\Theta \cdot \Theta \right)$
6:    **until** convergence
7:    **return** $\hat{\Theta}$
8: **end procedure**

Figure 4: Optimizing models for BPR with bootstrapping based stochastic gradient descent. With learning rate $\alpha$ and regularization $\lambda_\Theta$.

The two most common algorithms for gradient descent are either full or stochastic gradient descent. In the first case, in each step the full gradient over all training data is computed and then the model parameters are updated with the learning rate $\alpha$:

$$\Theta \leftarrow \Theta - \alpha \frac{\partial \text{BPR-OPT}}{\partial \Theta}$$

In general this approach leads to a descent in the 'correct' direction, but convergence is slow. As we have $O(|S||I|)$ training triples in $D_S$, computing the full gradient in each update step is not feasible. Furthermore, for optimizing BPR-OPT with full gradient descent also the skewness in the training pairs leads to poor convergence. Imagine an item $i$ that is often positive. Then we have many terms of the form $\hat{x}_{uij}$ in the loss because for many users $u$ the item $i$ is compared against all negative items $j$ (the dominating class). Thus the gradient for model parameters depending on $i$ would dominate largely the gradient. That means very small learning rates would have to be chosen. Secondly, regularization is difficult as the gradients differ much.

The other popular approach is stochastic gradient descent. In this case for each triple $(u,i,j) \in D_S$ an update is performed.

$$\Theta \leftarrow \Theta + \alpha \left( \frac{e^{-\hat{x}_{uij}}}{1 + e^{-\hat{x}_{uij}}} \cdot \frac{\partial}{\partial \Theta} \hat{x}_{uij} + \lambda_\Theta \Theta \right)$$

In general this is a good approach for our skew problem but the order in which the training pairs are traversed is crucial. A typical approach that traverses the data item-wise or user-wise will lead to poor convergence as there are so many consecutive updates on the same user-item pair – i.e. for one user-item pair $(u,i)$ there are many $j$ with $(u,i,j) \in D_S$.

To solve this issue we suggest to use a stochastic gradient descent algorithm that chooses the triples randomly (uniformly distributed). With this approach the chances to pick the same user-item combination in consecutive update steps is small. We suggest to use a bootstrap sampling approach with replacement because stopping can be performed at any step. Abandoning the idea of full cycles through the data is especially useful in our case as the number of examples is very large and for convergence often a fraction of a full cycle is sufficient. We choose the number of single steps in our evaluation linearly depending on the number of observed positive feedback $S$.

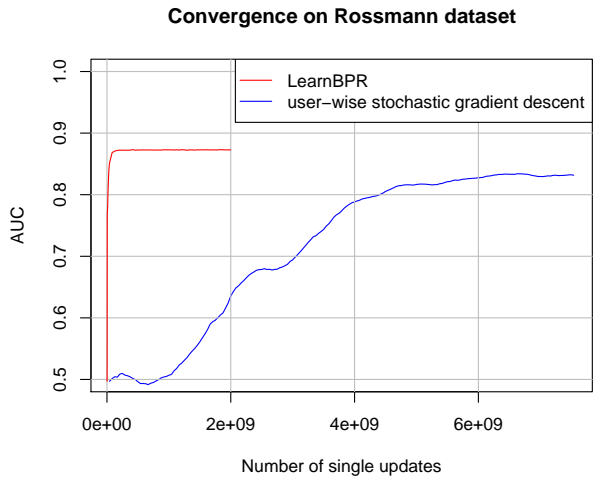

Figure 5: Empirical comparison of the convergence of typical user-wise stochastic gradient descent to our LEARNBPR algorithm with bootstrap sampling.

Figure 5 shows a comparison[1] of a typical user-wise stochastic gradient descent to our approach LEARNBPR with bootstrapping. The model is BPR-MF with 16 dimensions. As you can see LEARNBPR converges much faster than user-wise gradient descent.

### 4.3 Learning models with BPR

In the following we describe two state-of-the-art model classes for item recommendation and how we can learn them with our proposed BPR methods. We have chosen the two diverse model classes of matrix factorization [5, 12] and learned k-nearest-neighbor [8]. Both classes try to model the hidden preferences of a user on an item. Their prediction is a real number $\hat{x}_{ul}$ per user-item-pair $(u, l)$.

Because in our optimization we have triples $(u, i, j) \in D_S$, we first decompose the estimator $\hat{x}_{uij}$ and define it as:

$$\hat{x}_{uij} := \hat{x}_{ui} - \hat{x}_{uj}$$

[1]Details about the dataset and evaluation method can be found in Section 6.

Now we can apply any standard collaborative filtering model that predicts $\hat{x}_{ul}$.

It is important to note that even though we use the same models as in other work, we optimize them against another criterion. This will lead to a better ranking because our criterion is optimal for the ranking task. Our criterion does not try to regress a single predictor $\hat{x}_{ul}$ to a single number but instead tries to classify the difference of two predictions $\hat{x}_{ui} - \hat{x}_{uj}$.

#### 4.3.1 Matrix Factorization

The problem of predicting $\hat{x}_{ui}$ can be seen as the task of estimating a matrix $X : U \times I$. With matrix factorization the target matrix $X$ is approximated by the matrix product of two low-rank matrices $W : |U| \times k$ and $H : |I| \times k$:

$$\hat{X} := WH^t$$

where $k$ is the dimensionality/rank of the approximation. Each row $w_u$ in $W$ can be seen as a feature vector describing a user $u$ and similarly each row $h_i$ of $H$ describes an item $i$. Thus the prediction formula can also be written as:

$$\hat{x}_{ui} = \langle w_u, h_i \rangle = \sum_{f=1}^{k} w_{uf} \cdot h_{if}$$

Besides the dot product $\langle \cdot, \cdot \rangle$ in general any kernel can be used like in [11]. The model parameters for matrix factorization are $\Theta = (W, H)$. The model parameters can also be seen as latent variables, modeling the non-observed taste of a user and the non-observed properties of an item.

In general the best approximation of $\hat{X}$ to $X$ with respect to least-square is achieved by the singular value decomposition (SVD). For machine learning tasks, it is known that SVD overfits and therefore many other matrix factorization methods have been proposed, including regularized least square optimization, non-negative factorization, maximum margin factorization, etc.

For the task of ranking, i.e. estimating whether a user prefers one item over another, a better approach is to optimize against the BPR-OPT criterion. This can be achieved by using our proposed algorithm LEARNBPR. As stated before for optimizing with LEARNBPR, only the gradient of $\hat{x}_{uij}$ with respect to every model parameter $\theta$ has to be known. For the matrix factorization model the derivatives are:

$$\frac{\partial}{\partial \theta}\hat{x}_{uij} = \begin{cases} (h_{if} - h_{jf}) & \text{if } \theta = w_{uf}, \\ w_{uf} & \text{if } \theta = h_{if}, \\ -w_{uf} & \text{if } \theta = h_{jf}, \\ 0 & \text{else} \end{cases}$$

Furthermore, we use three regularization constants: one $\lambda_W$ for the user features $W$; for the item features $H$ we have two regularization constants, $\lambda_{H^+}$ that is used for positive updates on $h_{if}$, and $\lambda_{H^-}$ for negative updates on $h_{jf}$.

#### 4.3.2 Adaptive k-Nearest-Neighbor

Nearest-neighbor methods are very popular in collaborative filtering. They rely on a similarity measure between either items (item-based) or users (user-based). In the following we describe item-based methods as they usually provide better results, but user-based methods work analogously. The idea is that the prediction for a user $u$ and an item $i$ depends on the similarity of $i$ to all other items the user has seen in the past – i.e. $I_u^+$. Often only the $k$ most similar items of $I_u^+$ are regarded – the k-nearest neighbors. If the similarities between items are chosen carefully, one can also compare to all items in $I_u^+$. For item prediction the model of item-based k-nearest-neighbor is:

$$\hat{x}_{ui} = \sum_{l \in I_u^+ \land l \neq i} c_{il}$$

where $C : I \times I$ is the symmetric item-correlation/ item-similarity matrix. Hence the model parameters of kNN are $\Theta = C$.

The common approach for choosing $C$ is by applying a heuristic similarity measure, e.g. cosine vector similarity:

$$c_{i,j}^{\text{cosine}} := \frac{|U_i^+ \cap U_j^+|}{\sqrt{|U_i^+| \cdot |U_j^+|}}$$

A better strategy is to adapt the similarity measure $C$ to the problem by learning it. This can be either done by using $C$ directly as model parameters or if the number of items is too large, one can learn a factorization $HH^t$ of $C$ with $H : I \times k$. In the following and also in our evaluation we use the first approach of learning $C$ directly without factorizing it.

Again for optimizing the kNN model for ranking, we apply the BPR optimization criterion and use the LEARNBPR algorithm. For applying the algorithm, the gradient of $\hat{x}_{uij}$ with respect to the model parameters $C$ is:

$$\frac{\partial}{\partial \theta} \hat{x}_{uij} = \begin{cases} +1 & \text{if } \theta \in \{c_{il}, c_{li}\} \land l \in I_u^+ \land l \neq i, \\ -1 & \text{if } \theta \in \{c_{jl}, c_{lj}\} \land l \in I_u^+ \land l \neq j, \\ 0 & \text{else} \end{cases}$$

We have two regularization constants, $\lambda_+$ for updates on $c_{il}$, and $\lambda_-$ for updates on $c_{jl}$.

## 5 Relations to other methods

We discuss the relations of our proposed methods for ranking to two further item prediction models.

### 5.1 Weighted Regularized Matrix Factorization (WR-MF)

Both Pan et al. [10] and Hu et al. [5] have presented a matrix factorization method for item prediction from implicit feedback. Thus the model class is the same as we described in Section 4.3.1, i.e. $\hat{X} := WH^t$ with the matrices $W : |U| \times k$ and $H : |U| \times k$. The optimization criterion and learning method differ substantially from our approach. Their method is an adaption of a SVD, which minimizes the square-loss. Their extensions are regularization to prevent overfitting and weights in the error function to increase the impact of positive feedback. In total their optimization criterion is:

$$\sum_{u \in U} \sum_{i \in I} c_{ui}(\langle w_u, h_i \rangle - 1)^2 + \lambda ||W||_f^2 + \lambda ||H||_f^2$$

where $c_{ui}$ are not model parameters but apriori given weights for each tuple $(u, i)$. Hu et al. have additional data to estimate $c_{ui}$ for positive feedback and they set $c_{ui} = 1$ for the rest. Pan et al. suggest to set $c_{ui} = 1$ for positive feedback and choose lower constants for the rest.

First of all, it is obvious that this optimization is on instance level (one item) instead of pair level (two items) as BPR. Apart from this, their optimization is a least-square which is known to correspond to the MLE for normally distributed random variables. However, the task of item prediction is actually not a regression (quantitative), but a classification (qualitative) one, so the logistic optimization is more appropriate.

A strong point of WR-MF is that it can be learned in $O(\text{iter}(|S| k^2 + k^3 (|I| + |U|)))$ provided that $c_{ui}$ is constant for non-positive pairs. Our evaluation indicates that LEARNBPR usually converges after a subsample of $m \cdot |S|$ single update steps even though there are much more triples to learn from.

### 5.2 Maximum Margin Matrix Factorization (MMMF)

Weimer et al. [15] use the maximum margin matrix factorization method (MMMF) for ordinal ranking. Their MMMF is designed for scenarios with explicit feedback in terms of ratings. Even though their ranking MMMF is not intended for implicit feedback datasets, one could apply it in our scenario by giving all non-observed items the 'rating' 0 and the observed ones a 1 (see Figure 1). With these modifications their optimization criterion to be minimized would be quite similar to BPR applied for matrix factorization:

$$\sum_{(u,i,j) \in D_s} \max(0, 1 - \langle w_u, h_i - h_j \rangle) + \lambda_w ||W||_f^2 + \lambda_h ||H||_f^2$$

One difference is that the error functions differ – our hinge loss is smooth and motivated by the MLE. Additionally, our BPR-OPT criterion is generic and can be applied to several models, whereas their method is specific for MF.

Besides this, their learning method for MMMF differs from our generic approach LEARNBPR. Their learning method is designed to work with sparse explicit data, i.e. they assume that there are many missing values and thus they assume to have much less pairs than in an implicit setting. But when their learning method is applied to implicit feedback datasets, the data has to be densified like described above and the number of training pairs $D_S$ is in $O(|S||I|)$. Our method LEARNBPR can handle this situation by bootstrapping from $D_S$ (see Section 4.2).

## 6 Evaluation

In our evaluation we compare learning with BPR to other learning approaches. We have chosen the two popular model classes of matrix factorization (MF) and k-nearest-neighbor (kNN). MF models are known to outperform [12] many other models including the Bayesian models URP [9] and PLSA [4] for the related task of collaborative rating prediction. In our evaluation, the matrix factorization models are learned by three different methods, i.e. SVD-MF, WR-MF [5, 10] and our BPR-MF. For kNN, we compare cosine vector similarity (Cosine-kNN) to a model that has been optimized using our BPR method (BPR-kNN). Additionally, we report results for the baseline most-popular, that weights each item user-independent, e.g.: $\hat{x}_{ui}^{\text{most-pop}} := |U_i^+|$. Furthermore, we give the theoretical upper bound on AUC (np$_{\max}$) for any non-personalized ranking method.

### 6.1 Datasets

We use two datasets of two different applications. The *Rossmann* dataset is from an online shop. It contains the buying history of 10,000 users on 4000 items. In total 426,612 purchases are recorded. The task is to predict a personalized list of the items the user wants to buy next. The second dataset is the DVD rental dataset of *Netflix*. This dataset contains the rating behavior of users, where a user provides explicit ratings 1 to 5 stars for some movies. As we want to solve an implicit feedback task, we removed the rating scores from the dataset. Now the task is to predict if a user is likely to rate a movie. Again we are interested in a personalized ranked list starting with the movie that is most likely to be rated. For Netflix we have created a subsample of 10,000 users, 5000 items containing 565,738 rating actions. We draw the subsample such that every user has at least 10 items ($\forall u \in U : |I_u^+| \geq 10$) and each item has at least 10 users: $\forall i \in I : |U_i^+| \geq 10$.

### 6.2 Evaluation Methodology

We use the leave one out evaluation scheme, where we remove for each user randomly one action (one user-item pair) from his history, i.e. we remove one entry from $I_u^+$ per user $u$. This results in a disjoint train set $S_{\text{train}}$ and test set $S_{\text{test}}$. The models are then learned on $S_{\text{train}}$ and their predicted personalized ranking is evaluated on the test set $S_{\text{test}}$ by the average AUC statistic:

$$\text{AUC} = \frac{1}{|U|} \sum_u \frac{1}{|E(u)|} \sum_{(i,j) \in E(u)} \delta(\hat{x}_{ui} > \hat{x}_{uj}) \quad (2)$$

where the evaluation pairs per user $u$ are:

$$E(u) := \{(i,j) | (u,i) \in S_{\text{test}} \wedge (u,j) \notin (S_{\text{test}} \cup S_{\text{train}})\}$$

A higher value of the AUC indicates a better quality. The trivial AUC of a random guess method is 0.5 and the best achievable quality is 1.

We repeated all experiments 10 times by drawing new train/test splits in each round. The hyperparameters for all methods are optimized via grid search in the first round and afterwards are kept constant in the remaining 9 repetitions.

### 6.3 Results and Discussion

Figure 6 shows the AUC quality of all models on the two datasets. First of all, you can see that the two BPR optimized methods outperform all other methods in prediction quality. Comparing the same models among each other one can see the importance of the optimization method. For example all MF methods (SVD-MF, WR-MF and BPR-MF) share exactly the same model, but their prediction quality differs a lot. Even though SVD-MF is known to yield the best fit on the training data with respect to element-wise least square, it is a poor prediction method for machine learning tasks as it results in overfitting. This can be seen as the quality of SVD-MF decreases with an increasing number of dimensions. WR-MF is a more successful learning method for the task of ranking. Due to regularization its performance does not drop but steadily rises with an increasing number of dimensions. But BPR-MF outperforms WR-MF clearly for the task of ranking on both datasets. For example on Netflix a MF model with 8 dimensions optimized by BPR-MF achieves comparable quality as a MF model with 128 dimensions optimized by WR-MF.

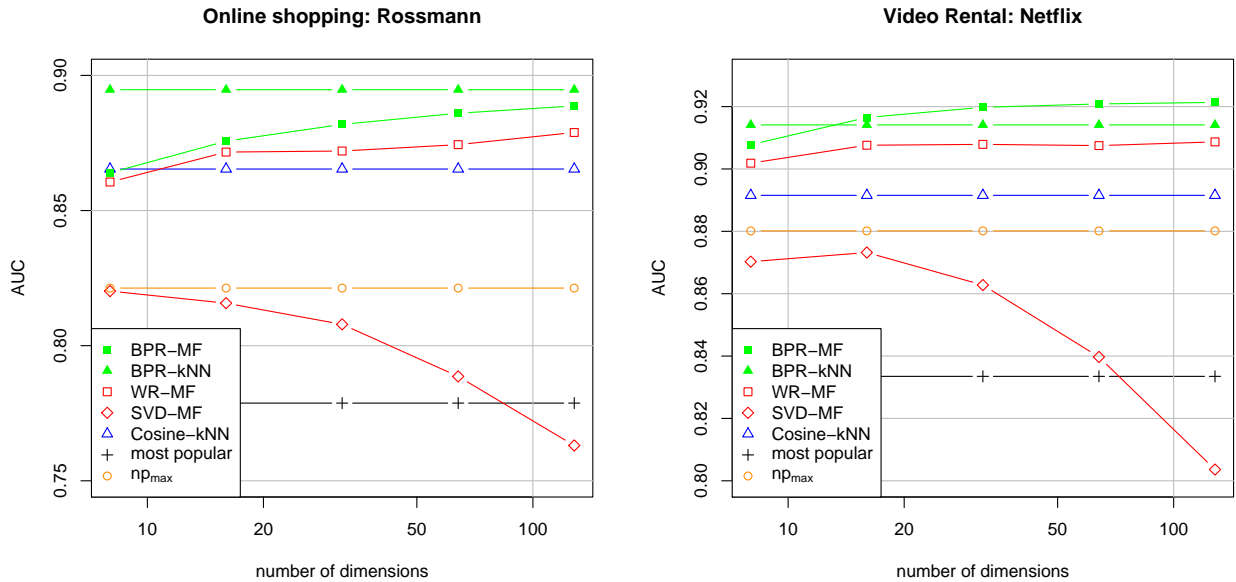

Figure 6: Area under the ROC curve (AUC) prediction quality for the Rossmann dataset and a Netflix subsample. Our BPR optimizations for matrix factorization BPR-MF and k-nearest neighbor BPR-kNN are compared against weighted regularized matrix factorization (WR-MF) [5, 10], singular value decomposition (SVD-MF), k-nearest neighbor (Cosine-kNN) [2] and the most-popular model. For the factorization methods BPR-MF, WR-MF and SVD-MF, the model dimensions are increased from 8 to 128 dimensions. Finally, $np_{max}$ is the theoretical upper bound for any non-personalized ranking method.

To summarize, our results show the importance of optimizing model parameters to the right criterion. The empirical results indicate that our BPR-OPT criterion learned by LEARNBPR outperforms the other state-of-the-art methods for personalized ranking from implicit feedback. The results are justified by the analysis of the problem (section 3.2) and by the theoretical derivation of BPR-OPT from the MLE.

### 6.4 Non-personalized ranking

Finally, we compare the AUC quality of our personalized ranking methods to the best possible non-personalized ranking method. In contrast to our personalized ranking methods, a non-personalized ranking method creates the same ranking $>$ for all users. We compute the theoretical upper-bound $np_{max}$ for any non-personalized ranking method by optimizing the ranking $>$ on the test set $S_{test}$[2]. Figure 6 shows that even simple personalized methods like Cosine-kNN outperform the upper-bound $np_{max}$ — and thus also all non-personalized methods — largely.

## 7 Conclusion

In this paper we have presented a generic optimization criterion and learning algorithm for personalized ranking. The optimization criterion BPR-OPT is the maximum posterior estimator that is derived from a Bayesian analysis of the problem. For learning models with respect to BPR-OPT we have presented the generic learning algorithm LEARNBPR that is based on stochastic gradient descent with bootstrap sampling. We have demonstrated how this generic method can be applied to the two state-of-the-art recommender models of matrix factorization and adaptive kNN. In our evaluation we show empirically that for the task of personalized ranking, models learned by BPR outperform the same models that are optimized with respect to other criteria. Our results show that the prediction quality does not only depend on the model but also largely on the optimization crite-rion. Both our theoretical and empirical results indicate that the BPR optimization method is the right choice for the important task of personalized ranking.

---

[2]We computed a real upper-bound but non-tight estimate on the AUC score. Please note that ranking by most-popular on test is not an upper bound on AUC. But in our experiments both AUC scores are quite similar, e.g. on Netflix with most-popular on test 0.8794 vs. our upper bound of 0.8801.

**Acknowledgements**

The authors gratefully acknowledge the partial co-funding of their work through the European Commission FP7 project MyMedia (www.mymediaproject.org) under the grant agreement no. 215006. For your inquiries please contact info@mymediaproject.org.

# References

[1] C. Burges, T. Shaked, E. Renshaw, A. Lazier, M. Deeds, N. Hamilton, and G. Hullender. Learning to rank using gradient descent. In *ICML '05: Proceedings of the 22nd international conference on Machine learning*, pages 89–96, New York, NY, USA, 2005. ACM Press.

[2] M. Deshpande and G. Karypis. Item-based top-n recommendation algorithms. *ACM Transactions on Information Systems. Springer-Verlag*, 22/1, 2004.

[3] A. Herschtal and B. Raskutti. Optimising area under the roc curve using gradient descent. In *ICML '04: Proceedings of the twenty-first international conference on Machine learning*, page 49, New York, NY, USA, 2004. ACM.

[4] T. Hofmann. Latent semantic models for collaborative filtering. *ACM Trans. Inf. Syst.*, 22(1):89–115, 2004.

[5] Y. Hu, Y. Koren, and C. Volinsky. Collaborative filtering for implicit feedback datasets. In *IEEE International Conference on Data Mining (ICDM 2008)*, pages 263–272, 2008.

[6] J. Huang, C. Guestrin, and L. Guibas. Efficient inference for distributions on permutations. In J. Platt, D. Koller, Y. Singer, and S. Roweis, editors, *Advances in Neural Information Processing Systems 20*, pages 697–704, Cambridge, MA, 2008. MIT Press.

[7] R. Kondor, A. Howard, and T. Jebara. Multi-object tracking with representations of the symmetric group. In *Proceedings of the Eleventh International Conference on Artificial Intelligence and Statistics, San Juan, Puerto Rico*, March 2007.

[8] Y. Koren. Factorization meets the neighborhood: a multifaceted collaborative filtering model. In *KDD '08: Proceeding of the 14th ACM SIGKDD international conference on Knowledge discovery and data mining*, pages 426–434, New York, NY, USA, 2008. ACM.

[9] B. Marlin. Modeling user rating profiles for collaborative filtering. In S. Thrun, L. Saul, and B. Schölkopf, editors, *Advances in Neural Information Processing Systems 16*, Cambridge, MA, 2004. MIT Press.

[10] R. Pan, Y. Zhou, B. Cao, N. N. Liu, R. M. Lukose, M. Scholz, and Q. Yang. One-class collaborative filtering. In *IEEE International Conference on Data Mining (ICDM 2008)*, pages 502–511, 2008.

[11] S. Rendle and L. Schmidt-Thieme. Online-updating regularized kernel matrix factorization models for large-scale recommender systems. In *RecSys '08: Proceedings of the 2008 ACM conference on Recommender systems*. ACM, 2008.

[12] J. D. M. Rennie and N. Srebro. Fast maximum margin matrix factorization for collaborative prediction. In *ICML '05: Proceedings of the 22nd international conference on Machine learning*, pages 713–719, New York, NY, USA, 2005. ACM.

[13] B. Sarwar, G. Karypis, J. Konstan, and J. Riedl. Incremental singular value decomposition algorithms for highly scalable recommender systems. In *Proceedings of the 5th International Conference in Computers and Information Technology*, 2002.

[14] L. Schmidt-Thieme. Compound classification models for recommender systems. In *IEEE International Conference on Data Mining (ICDM 2005)*, pages 378–385, 2005.

[15] M. Weimer, A. Karatzoglou, and A. Smola. Improving maximum margin matrix factorization. *Machine Learning*, 72(3):263–276, 2008.